\documentclass[acmlarge,manuscript,nonacm]{acmart}

\usepackage{xcolor}
\usepackage{caption}
\usepackage{subcaption}
\AtBeginDocument{%
  }

\setcopyright{acmlicensed}
\copyrightyear{2018}
\acmYear{2018}
\acmDOI{XXXXXXX.XXXXXXX}

\begin{document}

\title[LLM- versus Intent-based Chatbots]{Conversational Assistants in Knowledge-Intensive Contexts: Interactions with LLM- versus Intent-based Systems}

\author{S. Kernan Freire}
\orcid{0000-0001-8684-0585}
\affiliation{%
  \institution{Delft University of Technology}
  \streetaddress{Landbergstraat 15}
  \city{Delft}
  \postcode{2628 CE}
  \country{The Netherlands}}
\email{s.kernanfreire@tudelft.nl}

\author{C. Wang}
\orcid{0000-0001-8213-6582}
\affiliation{%
  \institution{Delft University of Technology}
  \streetaddress{Landbergstraat 15}
  \city{Delft}
  \postcode{2628 CE}
  \country{The Netherlands}}
\email{c.wang-16@tudelft.nl}

\author{E. Niforatos}
\orcid{0000-0002-0484-4214}
\affiliation{%
  \institution{Delft University of Technology}
  \streetaddress{Landbergstraat 15}
  \city{Delft}
  \country{Netherlands}}
\email{e.niforatos@tudelft.nl}

\renewcommand{\shortauthors}{Kernan Freire et al.}

\begin{abstract}
Conversational Assistants (CA) are increasingly supporting human workers in knowledge management. Traditionally, CAs respond in specific ways to predefined user intents and conversation patterns. However, this rigidness does not handle the diversity of natural language well. Recent advances in natural language processing, namely Large Language Models (LLMs), enable CAs to converse in a more flexible, human-like manner, extracting relevant information from texts and capturing information from expert humans but introducing new challenges such as ``hallucinations''. To assess the potential of using LLMs for knowledge management tasks, we conducted a user study comparing an LLM-based CA to an intent-based system regarding interaction efficiency, user experience, workload, and usability. This revealed that LLM-based CAs exhibited better user experience, task completion rate, usability, and perceived performance than intent-based systems, suggesting that switching NLP techniques can be beneficial in the context of knowledge management.
\end{abstract}



\maketitle

\section{Introduction}
In workplaces, Conversational Assistants (CAs) are increasingly used to support decision-making and knowledge management. Knowledge Management (KM) is a vital discipline that involves creating, sharing, and using an organization's knowledge. Integrated into a Knowledge Management System (KMS), CAs can be used to capture and share knowledge among workers through conversational interactions~\cite{KernanFreire2022ACognitive}.

The prevailing NLP technique for CAs in KM is intent-based~\cite{Wellsandt.2020}. Intent-based systems offer designers a high level of control over possible user interactions and can deliver consistent responses. However, the rigidity of this technique can result in frequent conversation breakdowns~\cite{rahman2017prog}. From the developer's perspective, intent-based systems are resource-intensive to create and maintain, requiring explicit definitions of all user intents, assistant responses, and examples of conversations~\cite{misargopoulos2022building}. On the other hand, LLM-based systems appear quick to deploy, and their superior NLP capabilities can be used to develop more robust conversational interactions.

A CA powered by a foundational Large Language Model (LLM) can answer general knowledge questions, yet they lack the specialized, context-specific knowledge of a workplace~\cite{bang2023multitask, zhao2023survey}. The information contained in the foundation models can be extended by providing context material, a process called Retrieval Augmented Generation (RAG)~\cite{lewis2020retrieval}. This involves an additional step in the response that retrieves context material relevant to the user's query from a data source (e.g., a local database or the web). 
However, LLM-based systems could introduce new problems due to their unpredictability. While the potential benefits and risks of LLM-based systems have been demonstrated by prior work~\cite{singhalLLMnature2023c,pan2024ieee,hu2024surveyknowledgeplm}, they have not been directly measured against intent-based systems in knowledge management interactions. As such, organizations and practitioners lack concrete evidence to decide whether to implement LLM- or intent-based CAs. Existing benchmarks typically assess question-answers pairs, such as the KILT benchmark for knowledge-intensive language tasks~\cite{petroniKILT2021}, but therefore do not capture the user interaction challenges associated with multi-turn queries in a larger conversation.

To assess the user interaction capabilities of LLM-based systems compared to intent-based systems in the context of KM, we conducted a between-groups lab study to compare them quantitatively in interaction efficiency, perceived workload, user experience, and usability while also qualitatively analyzing user feedback. The study participants were instructed to perform eight knowledge management tasks. As such, \textbf{this study contributes a deeper understanding of how switching to an LLM-based NLP system affects user interactions in the context of KM, where users are expected to retrieve, rate and share knowledge.}

\section{Background}
In the following section, we outline existing literature and the motivation for this work. We address advancements in CAs for KM, NLP techniques for CAs, and using LLMs for KM.

\subsection{Conversational assistants for knowledge management}
In the workplace, AI has evolved from automating simple tasks (e.g., customer support chatbots) to aiding complex decision-making (e.g., CAs for production line operators).
CA systems are being developed to improve the efficiency of knowledge capture and sharing. For example, \citet{fenoglio_tacit_2022} introduced a role-playing game for knowledge capture, and \citet{balayn2022ready} developed a game to elicit tacit knowledge. \citet{soliman_model_2020} proposed a knowledge retention model, while \citet{hoerner_using_2022} focused on troubleshooting support. Workers supported by CAs are emerging as socio-technical systems integrating user goals, tasks, and technology, adapting to varying environments and user needs~\cite{KernanFreire2022ACognitive, maedche2019ai, workshop2016}.

\subsection{Natural language processing for conversational assistants}

Intent-based systems, grounded in the principles of symbolic AI and rule-based processing, currently underpin most CAs. These systems, as \citet{stolcke2000dialogue} and \citet{luo2022critical} discuss, are designed to recognize and interpret user intents through predefined patterns and commands. They rely on learning from, for example, conversational patterns and rules, enabling them to respond to specific queries or execute tasks according to recognized user utterances. While limited in flexibility and adaptability, this approach provides a high degree of control and predictability in interactions. However, as \citet{rahman2017prog} noted, the inherent rigidity of intent-based systems presents substantial challenges. These include difficulties comprehending nuanced expressions and operating effectively in dynamic contexts where user needs and environmental factors may rapidly shift. In turn, this may hurt UX as intent-based systems often struggle to accommodate atypical conversation patterns and phrasing that a human could easily understand~\cite{rahman2017prog}. Indeed, the literature emphasizes how intent-based chatbots frequently fail to help users achieve their goals~\cite{folstad2021future,folstad2017chatbots,lee2021exploring,meyer2021sorry}.

In contrast to the rigid intent-based systems, LLMs are flexible and nuanced~\cite{mahmood2023llmpowered}. LLMs such as Gemini family\footnote{\url{https://deepmind.google/technologies/gemini}--last accessed \today.}, Claude family\footnote{\url{https://www.anthropic.com/news/introducing-claude}--last accessed \today.}, or GPT-4 family~\cite{openai2023gpt4} can understand, generate, and interact with human language at a sophisticated level. In text generation, LLMs mark a significant step forward from their predecessors, Recurrent Neural Networks, largely due to their attention to the surrounding context when processing text~\cite{zhao2023survey}. This innate ability of LLMs supports more intelligent and adaptive interactions than intent-based systems at the risk of generating ``hallucinations'', seemingly plausible but inaccurate responses~\cite{min2023recent,bang2023multitask, zhao2023survey}. 

\section{User study}
\subsection{Context and research question}
Operating production lines is an ideal context for investigating CAs for KM, as the work is knowledge-intensive, dynamic, and fast-paced. Factories offer a rich resource for NLP with extensive text documents, such as work instructions and machine manuals, and continuous knowledge creation by discovering new machine setups and problem solutions for the ever-changing conditions. Yet, the unstructured text and inconsistent terminology, typically encountered in manufacturing contexts, can be challenging for NLP~\cite{edwards2008clustering}. State-of-the-art NLP, such as LLMs, can mitigate these challenges. For example, they can effectively harness the information in these documents using RAG\cite{lewis2020retrieval}, which previously required manual retrieval.

As researchers in a large innovation and research project, we witnessed a shift in the perceived usefulness of the CAs at the factories when we integrated LLMs in early 2023. We integrated LLMs for RAG and capturing and sharing knowledge with workers in a more flexible way than the techniques available to intent-based CAs, such as FAQs and form-filling. However, we wanted to empirically confirm our observations, formulating the following research question: \textbf{How do LLM- and intent-based conversational assistants compare in interaction efficiency, system usability, user experience, and perceived workload for workers?}\label{RQ1}

\subsection{Systems: LLM and intent-based conversational assistants}
For this study, we developed and evaluated two conversational assistants (CAs) with the same functionality but different NLP techniques: intent- and LLM-based. The CAs were developed with two factories to support workers through conversational knowledge sharing and information retrieval. The capturing component of knowledge sharing involves recording how workers solve issues or set up the machines for a product, whereas knowledge sharing involves delivering it to other workers when they request it. The factory management had observed significant disparities between worker shift performance, poor adherence to standard working procedures, and inefficient knowledge sharing practices. Despite the top-down initiation of the project, we involved the workers throughout the design \& development process to ensure that we also met their needs and values (participatory design~\cite{bjorgvinsson2010participatory}). This consisted of multiple semi-structured interviews, observations, and user evaluations. The systems had the following capabilities: capturing and sharing product-specific machine settings and advice; capturing and sharing issue-handling knowledge; FAQs with answers prepared by experts; and information retrieval from standard work instructions.

\begin{figure}[ht]
    \begin{subfigure}{0.45\columnwidth}
        \includegraphics[width=0.8\columnwidth]{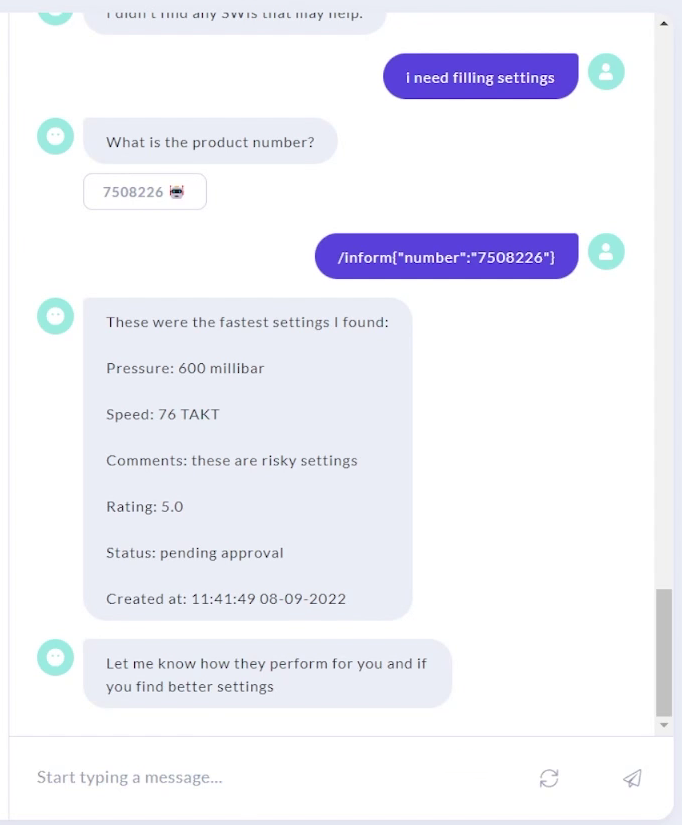}
        \caption{}
        \Description{TBD}
        \label{fig:intent-ui}
    \end{subfigure}
    \begin{subfigure}{0.45\columnwidth}
        \includegraphics[width=0.8\columnwidth]{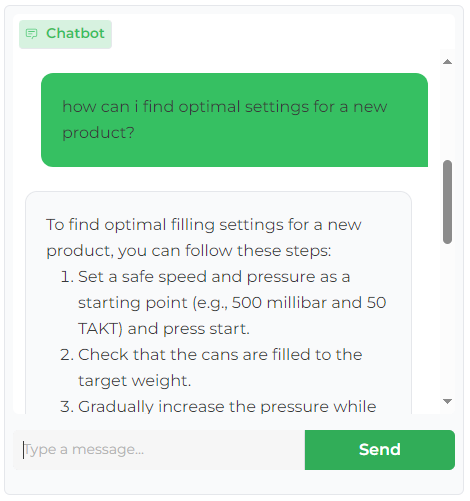}
        \caption{}
        \Description{TBD}
        \label{fig:llm-ui}
    \end{subfigure}
    \caption{(Conversational) User Interfaces (UIs) of (a) the Intent-based and (b) the LLM-based congitive assistants.}
    \label{fig:UIs}
\end{figure}

The \textbf{intent-based CA} relies on a list of user intents, and conversational patterns for which examples are provided. This data is used to train an intent classifier, entity extractor, and rules that define how the system should respond. Responses could include a text response to users, executing functions to insert information into a knowledge base, displaying a standard work instruction, or entering a form-filling loop\footnote{\url{https://rasa.com/docs/rasa/forms/}--last accessed \today.} to collect the necessary information from the user. Training phrases for each intent were used to train a model to classify the intent and extract desired named entities, such as machine component names. It was built using Rasa X\footnote{\url{https://legacy-docs-rasa-x.rasa.com/docs/rasa-x/1.0.x/}--last accessed \today.}, a popular conversational AI framework and featured a simple chat interface (see Figure~\ref{fig:intent-ui}).

Conversely, the \textbf{LLM-based system} was built using LlamaIndex for the backend \footnote{\url{https://docs.llamaindex.ai/}--last accessed \today.} and Gradio\footnote{\url{https://www.gradio.app/}--last accessed \today.} to build a comparable chat interface (see Figure~\ref{fig:llm-ui}, and the GPT-3.5 API (version: gpt-3.5-turbo-0613)\footnote{\url{https://platform.openai.com/docs/models/gpt-3-5}--last accessed \today.} for LLM calls. The behavior of the LLM-based system was defined using a system prompt and a context chat mode which retrieved relevant information from provided documents in every conversation turn, based on a technique called retrieval augmented generation (RAG)\footnote{\url{https://docs.llamaindex.ai/en/latest/examples/chat_engine/chat_engine_context.html}--last accessed \today.}. The word embeddings, which are used for finding relevant document sections to respond to the user's queries, were generated using the text-embedding-ada-002 model. We used the system prompt to instruct the LLM that it was an assistant for factory workers and to use the provided context material when responding. The context material contained relevant information and instructions that matched the knowledge base of the intent-based system in content.


\subsection{Mixed methods user study}
As a preliminary investigation into the differences in interaction efficiency and UX between the NLP techniques, we conducted a between-group user study with students.
First, participants were introduced to CAs and factory operations through a lecture and videos. After reading and signing the informed consent form, the participants were instructed to access the chat interface on their laptops or smartphones. Then, participants were instructed to complete eight information and knowledge sharing tasks assigned to them within ten minutes, as shown in Table~\ref{tab:tasklist}. The eight knowledge exchange tasks included six information retrieval tasks, a feedback task (Task 5), and a knowledge sharing task (Task 8). The most complex tasks are 6 and 8, which require providing several pieces of information; for example, to complete task 8 successfully, participants must specify the current product number, machine pressure and speed, and free-text comments. This preliminary study focuses on the CA's user interaction capabilities, therefore participants were not asked to act upon the retrieved information. After completing the tasks, participants completed a survey, starting with perceived workload (NASA-TLX) and ending with open feedback and demographics.

\begin{table}[ht]
\centering
\caption{Task List}
\label{tab:tasklist}
\begin{tabular}{|c|l|}
\hline
\textbf{Number} & \textbf{Task Description} \\ 
\hline
1 & Find instructions on how to perform a prerun \\
2 & Find instructions on how to prepare the weight checker \\
3 & Find instructions on how to start filling \\
4 & Find filling settings for product 7508226 \\
5 & Give a rating of 1 to the filling settings you were given \\
6 & Ask for help with this problem: ``Symptom: The filler is foaming. Error code: 33. Product: 7508226.'' \\
7 & Find instructions on how to find new filling settings \\
8 & Record the current filling machine settings \\
\hline
\end{tabular}
\end{table}

\subsubsection{Participants}
We recruited \textit{N=55} students to participate in the study. Although their educational background differs from factory workers, we believe their experience using chatbots is comparable to the dominant age group (17-29) is comparable. Most participants fell into the 17-29 bracket (\textit{n=47}), leaving two in the 30-39 bracket, and one preferred not to disclose. Gender was distributed as follows: \textit{n=26} women, \textit{n=21} men, two non-binary, and one did not disclose. We removed three cases for completing the tasks unrealistically fast (<60 seconds) and two cases for taking longer than 660 seconds. The participants were recruited in subsequent cohorts from the same masters-level course. Participants for the intent-based condition (\textit{n=17}) were collected in August 2022. Conversely, the data for the LLM condition (\textit{n=35}) was collected in November 2023.
  
\section{Results: LLM- versus intent-based conversational assistants}
Before selecting our statistical analysis methods, we conducted pre-tests, such as the Shapiro-Wilk tests for assessing data normality and Levene's tests to check the equality of variances. We used independent samples t-tests and Mann-Whitney U-tests for the parametric and non-parametric tests, respectively.

In assessing the effectiveness of the two groups — Intent and LLMs — in user experience, several dimensions were evaluated using the System Usability Scale (SUS), User Experience Questionnaire (UEQ), NASA-TLX, task time, and task completion rate. The results revealed distinct differences between the two groups (intent-based vs. LLM-based) across some measured facets, as presented below.

\begin{figure}[ht]
    \begin{subfigure}{0.3\columnwidth}
        \includegraphics[width=0.8\columnwidth]{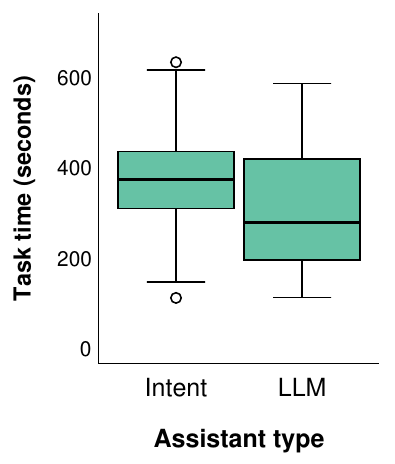}
        \caption{}
        \Description{TBD}
        \label{fig:task_time}
    \end{subfigure}
    \begin{subfigure}{0.3\columnwidth}
        \includegraphics[width=0.8\columnwidth]{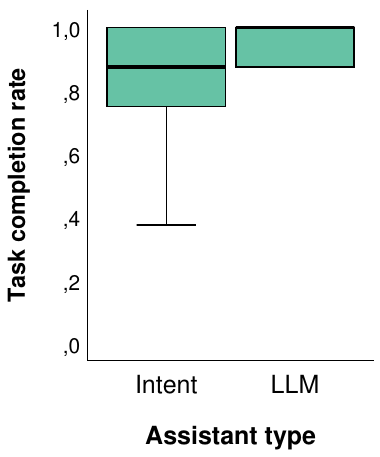}
        \caption{}
        \Description{TBD}
        \label{fig:taskCompletionRate}
    \end{subfigure}
    \begin{subfigure}{0.3\columnwidth}
        \includegraphics[width=0.8\columnwidth]{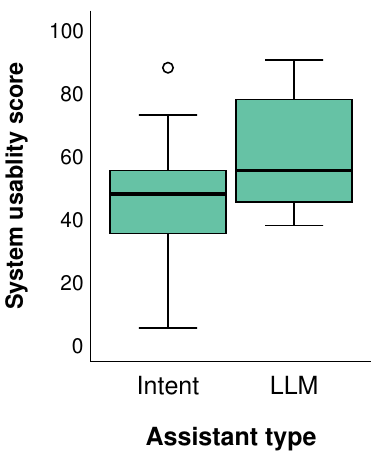}
        \caption{}
        \Description{TBD}
        \label{fig:SUS}
    \end{subfigure}
    \caption{Task time (a), Task completion rate* (b), and System usability score* (c) between the Intent and LLM groups}
    \label{fig:graphs}
\end{figure}

\subsection{Task performance}


Task time, which represents the time users spent completing tasks, was not significantly different between groups (\textit{t}(48) = 1.864, \textit{p} = .068)(see Figure~\ref{fig:task_time}). It was measured automatically by the time spent on the survey page containing the task instructions. The Intent group had a mean task time of 376.25 seconds (SD (Standard deviation) = 131.43), while the LLM group had a shorter mean time of 301.11 seconds (SD = 141.86).


That said, a significant difference was found in the task completion rate (see Figure~\ref{fig:taskCompletionRate}). The LLM group achieved a higher median task completion rate (1.00) compared to the Intent group (.88) (\textit{U} = 153.50, \textit{p} = .006). This suggests that users in the LLM condition were more successful in completing the assigned interactions.

\subsection{System usability}


For the SUS, the Intent group reported a mean score of 44.85 (SD = 16.75), while the LLM group demonstrated a higher mean score of 59.85 (SD = 17.47)(see Figure~\ref{fig:SUS}). This difference was statistically significant (\textit{t}(48) = -2.958, \textit{p} = .005), indicating that users found the LLM condition to be more usable compared to the Intent condition.

\subsection{User experience}

\begin{figure}[ht]
    \centering
    \includegraphics[width=0.90\columnwidth]{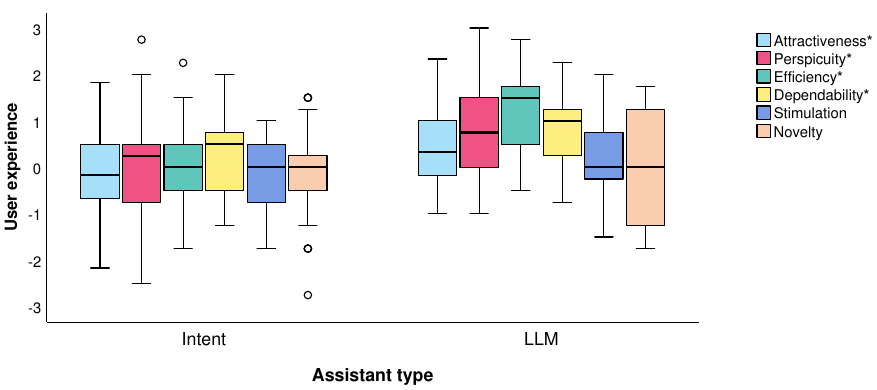}
    \caption{User Experience (UEQ scores) between the LLM and Intent groups}
    \Description{TBD}
    \label{fig:UX}
\end{figure}

We used the UEQ questionnaire to compare self-reported UX between the two groups (intent-based vs. LLM-based). These are reported as six dimensions (see Figure~\ref{fig:UX}). Regarding the Attractiveness, the Intent group's mean score was -.13 (SD = .85), compared to .46 (SD = .91) for the LLM condition. This difference was statistically significant (\textit{t}(48) = -2.25, \textit{p} = .029), suggesting that users perceived the LLM condition as more attractive. Regarding Perspicuity, the clarity of the user interface, the Intent group scored a median of .25, while the LLM group scored significantly higher with a median of .75, with a U-value of \textit{U} = 154.00 and p-value of \textit{p} = .009. This result indicates a clearer and more understandable interface in the LLM condition. Efficiency also showed a significant difference. The Intent group's mean score was .08 (SD = .87), whereas the LLM group scored a mean of 1.10 (SD = .91), resulting in a significant difference (\textit{t}(48) of -3.90 and \textit{p} < .001). This suggests that users found the LLM condition more efficient. When examining the Dependability aspect, the Intent group had a mean score of .25 (SD = .79), and the LLM group scored higher with a mean of .88 (SD = .85). This difference was significant (\textit{t}(48) = -2.54, \textit{p} = .01), indicating a greater sense of dependability perceived by users in the LLM condition. The remaining dimensions, stimulation, and novelty, did not significantly differ between groups, so we omitted the test details for brevity.



\subsection{Workload}

\begin{figure}[ht]
    \centering
    \includegraphics[width=0.90\columnwidth]{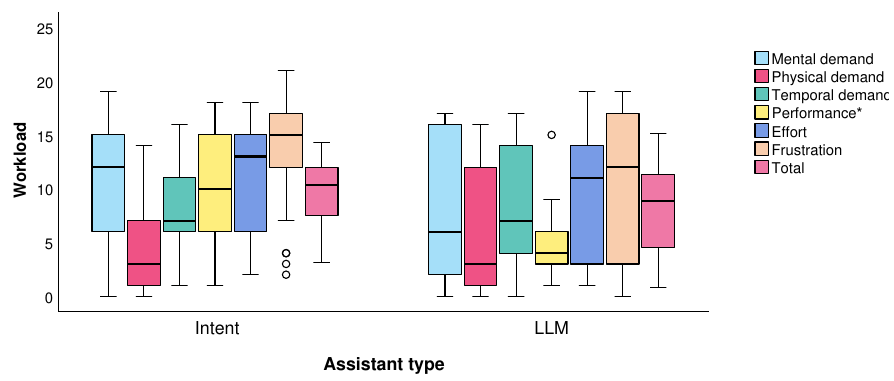}
    \caption{Workload (NASA-TLX) between the LLM and Intent conditions}
    \Description{TBD}
    \label{fig:workload}
\end{figure}

NASA-TLX is a 21-point scale where a high score indicates a high perceived workload. Of all the components, only ``performance'' showed a significant difference between the conditions (see Figure~\ref{fig:workload}). Namely, the LLM group had a significantly better median score of 4 compared to 10 for the Intent group (\textit{U} = 123.50, \textit{p} = .001). This indicates that users in the LLM condition experienced a higher sense of performance. For the mean perceived workload, the score for the Intent group was 9.70 (SD = 3.03), whereas it was 7.89 (SD = 4.54) for the LLM group. This was not significantly different, as shown by the following test scores: \textit{t}(48) = 1.49, \textit{p} = .15, suggesting both conditions experienced a similar overall workload.

\section{Discussion}
\subsection{Insights and implications for practice}

Our study suggests that LLM-based assistants could help workers perform knowledge management tasks more effectively than intent-based systems. Participants were more successful in completing the assigned tasks, which is also reflected in the perceived performance (NASA-TLX), user experience (UEQ), and usability (SUS). We believe the observed differences could be largely attributed to the flexibility and superior understanding of LLM-based conversational assistants (CAs) compared to intent-based ones. The following comments from the LLM condition support this view: ``It is reliable and trustworthy. Whatever instructions it gives me, I understood them well and believed it. (L38)''; and ``It felt quite intuitive, really easy to ask the `right' questions. (L56)''. Conversely, once the intent-based system went down the wrong `conversation path', it was more troublesome to recover from, sometimes requiring restarting the conversation, confirming existing literature~\cite{meyer2021sorry,folstad2021future,folstad2017chatbots,lee2021exploring}. Comments from the participants in the intent condition support these observations; for example, ``chatbot is not flexible, you have to follow its structure. (I29)'', and ``Annoying how you sometimes have to rephrase what you want. (I11)''. Interestingly, this did not lead to a difference in total interaction time between conditions, perhaps due to the sometimes slower and more verbose responses from the LLM CA. In contrast, the intent-based CA always responded concisely and, in some cases, provided buttons to suggest user responses.

Participants from the LLM condition noted a lack of input suggestions despite its superior NLP, as stated by participant L48: ``I would still like some suggestions or templates on how I can interact with.'' Furthermore, participants requested more efficient ways of interacting for frequently recurring patterns as L36 remarked: ``I think that actions that happen a lot (like, I assume, saving settings etc.) should not require typing so much on a smartphone screen''. Furthermore, multimodality will be important in knowledge-intensive contexts where visualizations are more applicable than text-based instructions, as mentioned by participant L37: ``Maybe consider to take the visualization tool to show the key information/parameter for a clearer understanding between the operator and the chatbox.''

Overall, the results demonstrate several advantages LLM-based CAs could bear over intent-based systems in terms of their abilities to exchange information with users and UX. However, it is important to consider that when intent-based systems fail, the worst that can happen is a misunderstanding, usually obvious to the users. Conversely, an LLM-based system could hallucinate an answer that appears plausible to the user. Assuming the chance of hallucinations will never be zero, it is crucial to consider the ethical, productivity, and safety implications. 

\subsection{Limitations and Future Work}
The participants were not factory workers by training, so we introduced them to the factory context before the study. Furthermore, we believe the results are valid as the experiment focused on interacting with the CA and did not require acting upon retrieved information. That being said, future work could consider the quality of the system's response and the impact on work. While an LLM-based system might be more successful in completing the interactions, the impact of ``hallucinations'' and other quality issues, such as wordy responses with irrelevant details, may negatively impact work considerably. Furthermore, in the context of knowledge sharing, future work could measure the quality of the captured knowledge.

A key challenge was comparing the two NLP systems fairly. For example, suggested responses are a characteristic feature of intent-based CA, but we decided against using them for the LLM condition as they are not commonly used. Still, future work could explore how the inclusion of suggested responses would affect the interactions of LLM-based CA. Considering there haven't been major improvements to intent or LLM-based technologies since their respective tests, we believe the two CAs we built still represent generic tools of their type.

Ultimately, we spent significantly more time developing the intent-based system to ensure it worked reliably for this study. Even so, the results revealed relatively low SUS scores, namely 45 for the intent condition and 60 for the LLM conditions, suggesting these tools do not fully represent the state-of-the-art. Building on this research, future work could develop a tool to support organizations choosing between different NLP techniques, such as intent- and LLM-based CAs. For example, on the one hand, including investments such as data, development, and user testing, and on the other hand, approximating the expected benefits, user experience, risks, and challenges.

\section{Conclusion}
LLM-based Conversational Assistants (CAs) are shown to improve interaction effectiveness over their intent-based counterparts for Knowledge Management (KM) tasks such as information retrieval, rating the retrieved information, and sharing new information. While the conversations with the intent-based CA often went down the wrong path and were difficult to recover from, the LLM-based CA's superior flexibility and intelligence in processing natural language enabled it to complete the KM interaction goals more successfully.

\begin{acks}

\end{acks}

\bibliographystyle{ACM-Reference-Format}
\bibliography{references}

\end{document}